\documentclass{elsart}
\usepackage[dvips]{graphicx}
\begin{document}
\begin{frontmatter}

\title{Search for $\beta^+$EC and ECEC processes in $^{112}$Sn and 
$\beta^-\beta^-$ decay 
of $^{124}$Sn to the excited states of $^{124}$Te}
\author[ITEP]{A.S.~Barabash$^{\;1)}$}\thanks{Corresponding author, 
Institute of Theoretical and Experimental Physics, B.~Cheremushkinskaya 25,
117218 Moscow, Russia,  e-mail: barabash@itep.ru,
tel.: 007 (495) 129-94-68, fax: 007 (495) 127-08-33},
\author[CENBG]{Ph.~Hubert},
\author[CENBG]{A.~Nachab},
\author[ITEP]{S.I.~Konovalov},
%\author[CENBG]{F.~Perrot},
\author[ITEP]{I.A.~Vanyushin},
\author[ITEP]{V.~Umatov}
\address[ITEP]{Institute of Theoretical and Experimental Physics, B.\
Cheremushkinskaya 25, 117218 Moscow, Russian Federation}
\address[CENBG]{Centre d'Etudes Nucl\'eaires,
IN2P3-CNRS et Universit\'e de Bordeaux, 33170 Gradignan, France}
%\submitted{Submitted to Nuclear Physics A }
\date{ }

\begin{abstract}

Limits on $\beta^+$EC and ECEC processes in $^{112}$Sn and on 
$\beta^-\beta^-$ decay 
of $^{124}$Sn to the excited states of $^{124}$Te have been obtained using 
a 380 cm$^3$ HPGe detector and an external source consisting of natural tin.
A limit with 90\% C.L. on the $^{112}$Sn half-life of 
$0.92\times 10^{20}$ y for the ECEC(0$\nu$) transition to the $0^+_3$ excited state 
in $^{112}$Cd (1871.0 keV) has been established. This transition is discussed in the context 
of a possible enhancement of the decay rate by several 
orders of magnitude given that the ECEC$(0\nu)$ process is nearly degenerate 
with an excited state in the daughter nuclide. Prospects for investigating 
such a process in future experiments are discussed.
The $\beta^-\beta^-$ decay limits for $^{124}$Sn to the excited states of $^{124}$Te 
were obtained on the level of $(0.8-1.2)\times 10^{21}$ y at the 90\% C.L.   


\end{abstract}
\end{frontmatter}

\newpage

\section{Introduction}
Interest in neutrinoless double-beta decay has seen a significant renewal in 
recent years after evidence for neutrino oscillations was obtained from the 
results of atmospheric, solar, reactor and accelerator  neutrino 
experiments (see, for example, the discussions in \cite{VAL06,BIL06,MOH06}). 
These results are impressive proof that neutrinos have a non-zero mass. However,
 the experiments studying neutrino oscillations are not sensitive to the nature
 of the neutrino mass (Dirac or Majorana) and provide no information on the 
absolute scale of the neutrino masses, since such experiments are sensitive 
only to the difference of the masses, $\Delta m^2$. The detection and study 
of $0\nu\beta\beta$ decay may clarify the following problems of neutrino 
physics (see discussions in \cite{PAS03,MOH05,PAS06}):
 (i) neutrino nature: 
whether the neutrino is a Dirac or a Majorana particle, (ii) absolute neutrino
 mass scale (a measurement or a limit on $m_1$), (iii) the type of neutrino 
mass hierarchy (normal, inverted, or quasidegenerate), (iv) CP violation in 
the lepton sector (measurement of the Majorana CP-violating phases).
At the present time only limits on the level of $\sim 10^{24} - 10^{25}$ y 
for half-lives and $\sim 0.3-1$ eV for effective Majorana neutrino 
mass $\left<m_\nu\right>$ have been obtained in the best modern experiments 
(see recent reviews \cite{BAR06,CRE06,AVI07}).     

In connection with the $0\nu\beta\beta$ decay, the detection of double
beta decay with emission of two neutrinos $(2\nu\beta\beta)$, which is
an allowed process of second order in the Standard Model, provides the
possibility for experimental determination  of the 
nuclear matrix  elements (NME) involved
in the double beta decay  processes.  This leads to the development of
theoretical  schemes for nuclear  matrix-element calculations  both in
connection   with  the   $2\nu\beta\beta$  decays   as  well   as  the
$0\nu\beta\beta$ decays (\cite{ROD06,KOR07,KOR07a}).  
At present,
$2\nu\beta\beta$  to  the ground  state  of the  final
daughter   nucleus  has been   measured  for   ten   nuclei 
(a review of results is given in
Ref.~\cite{BAR06a}).

Recently, it has been pointed out that the $2\nu\beta\beta$ decay allows 
one to investigate particle properties, in particular whether the Pauli 
exclusion principle is violated for neutrinos and thus neutrinos at least partially obey
Bose-Einstein statistics \cite{DOL05,BAR07b}. 

The $\beta\beta$  decay can proceed through transitions  to the ground
state as  well as to various  excited states of  the daughter nucleus.
Studies of  the latter transitions provide supplementary
information about  $\beta\beta$ decay.  

As it was shown in Ref.~\cite{BAR90}, by
using   low-background  facilities   utilizing  High Purity Germanium
(HPGe)   detectors,  the 
$2\nu\beta\beta$ decay  to the $0^+_1$  level in the  daughter nucleus
may  be detected. 
Soon after this double beta decay of $^{100}$Mo to 
the 0$^+$ excited state at 1130.29 keV in $^{100}$Ru was observed 
\cite{BAR95}. This result was confirmed in independent experiments 
 \cite{BAR99,DEB01,ARN07}. In 2004 for the first 
time the transition was detected in $^{150}$Nd 
\cite{BAR04}.
Recently additional isotopes, have become of interest in
studies of $2\nu\beta\beta$ decay to the $0^+_1$ level ~\cite{BAR00,BAR04a,BAR07}.

The $0\nu\beta\beta$ transition  to excited
states of  daughter nuclei has a clear signature  for 
such  decays. It is
worthy of a special  note here that in addition to two electrons with a fixed
total energy, there are one  ($0^+ - 2^+$ transition) or two ($0^+
-  0^+$ transition) photons with their energies  being strictly  fixed.  
In  a hypothetical experiment  which detects all the decay products  with a
high efficiency and a high  energy resolution the background can be
reduced to nearly zero.  It is possible that  this idea will be  used in
future experiments featuring a large mass of the isotope under study
(as it was mensioned in ~\cite{BAR04,BAR00,SUH00}). In
Ref.~\cite{SIM02} it  was stated that detection  of this transition
can distinguish  between the
$0\nu\beta\beta$   mechanisms (light and heavy Majorana neutrino
exchange mechanisms, trilinear R-parity breaking mechanisms etc.).
So   the   search  for   $\beta\beta$  
transitions to excited states has its own special interest.

Most double beta decay investigations have concentrated on the $\beta^-\beta^-$
 decay. Much less attention has been given to the investigation of 
$\beta^+\beta^+$, $\beta^+$EC and ECEC processes (here EC denotes electron 
capture). There are 34 candidates for these processes. Only six nuclei can undergo
 all of the above mentioned processes and 16 nuclei can undergo $\beta^+$EC 
and ECEC while 12 can undergo only ECEC. Detection of the neutrinoless mode in
 the above processes enable one to determine the effective Majorana neutrino 
mass $\left<m_\nu\right>$ and parameters of right-handed current admixture in 
electroweak interaction ($\left<\lambda\right>$ and $\left<\eta\right>$). 
Detection of the two-neutrino mode in the above processes lets one determine 
the magnitude of the nuclear matrix elements involved, which is very important 
in view of the theoretical calculations for both the $2\nu$ and the $0\nu$ 
modes of double beta decay. Interestingly, it was demonstrated in Ref. 
\cite{HIR94} that if the $\beta^-\beta^-(0\nu)$ decay is detected, then the
 experimental limits on the $\beta^+EC(0\nu)$ half-lives can be used to obtain
 information about the relative importance of the Majorana neutrino mass and
 right-handed current admixtures in electroweak interactions.

The $\beta^+\beta^+$ and $\beta^+$EC processes are less favorable due to 
smaller kinetic energy available for the emitted particles and
Coulomb barrier for the positrons. However, an 
attractive feature of these processes from the experimental point of view is
 the possibility of detecting either the coincidence signals from four (two)
 annihilation $\gamma$-rays and two (one) positrons, or the annihilation 
$\gamma$-rays only. It is difficult to investigate the ECEC process because one
 only detects the low energy X-rays. It is also interesting to search for 
transitions to the excited states of daughter nuclei, which are easier to detect
 given the cascade of higher energy gammas \cite{BAR94}. 
In Ref. \cite{WIN55} it was the first mentioned that in the case of ECEC(0$\nu$)
transition a resonance condition can exist for transition to the "right energy"
of the excited level for the daughter nucleus, here the decay energy is close to zero.
In 1982 the same idea was proposed for the transition to the ground state 
\cite{VOL82}. In 1983 this possibility was discussed for the transition 
$^{112}$Sn to $^{112}$Cd ($0^+$; 1871 keV) \cite{BER83}. In 2004 the idea 
was reanalyzed in Ref. \cite{SUJ04} and new resonance conditions for the 
decay were formulated. 
The possible enhancement of the transition rate was estimated as $\sim$ 10$^6$
 \cite{BER83,SUJ04}. This means that this process starts to be competitive with
 $0\nu\beta\beta$ decay for the neutrino mass sensitivity and is  
interesting to check experimentally. There are several candidates for 
which resonance transition, to the ground ($^{152}$Gd, $^{164}$Eu and $^{180}$W)
 and to the excited states ($^{74}$Se, $^{78}$Kr, $^{96}$Ru, $^{106}$Cd, 
$^{112}$Sn, $^{130}$Ba, $^{136}$Ce and $^{162}$Er)) of daughter nuclei exist 
\cite{BAR07,SUJ04}. The precision needed to realize resonance condition is well
 below 1 keV. To select the best candidate from the above list one will have 
to know the atomic mass difference with an accuracy better than 1 keV and
such measurements are planned for the future. Recently the experimental search
 for such a resonance transition in $^{74}$Se to $^{74}$Ge ($2^+$; 1206.9 keV) was
 performed yielding a limit $T_{1/2} > 5.5\times10^{18}$ y \cite{BAR07a}. Very
 recently $^{112}$Sn was investigated \cite{KIM07,DAW07} and a limit of 
$T_{1/2} > 1.6\times10^{18}$ y was obtained for the transition
to the $0^+$ state at 1871 keV \cite{DAW07}.

In this article the results of an experimental investigation of the $\beta^+$EC and ECEC 
processes in $^{112}$Sn and $\beta^-\beta^-$ decay 
of $^{124}$Sn to the excited states of $^{124}$Te are presented. 

\begin{figure}[h]
\begin{center}
\includegraphics[width=10.5cm]{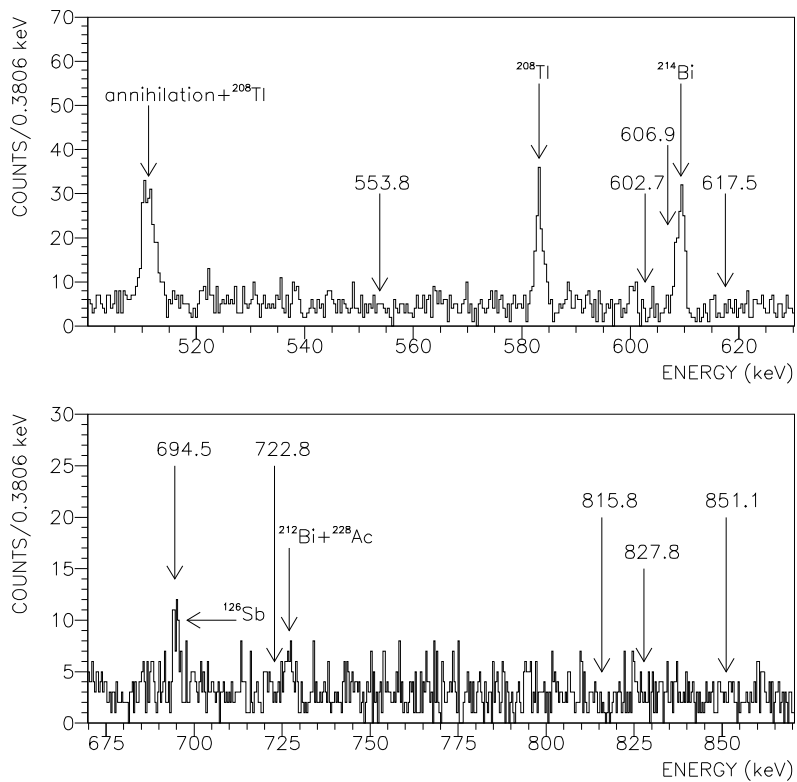}
\caption{Energy spectrum with 3975.3 g of natural Sn for 2385.4 h of measurement 
in the ranges investigated ([500-630] and [670-870] keV).}  
\label{fig_1}
\end{center}
\end{figure}

\section{Experimental}

The experiment was performed in the Modane Underground Laboratory 
at a depth of 4800 m w.e.. The natural tin sample, guaranteed to be 99.99 \% pure, 
was measured using a 380 cm$^3$ low-background HPGe detector.

The HPGe spectrometer is a p-type crystal with  
the cryostat, endcap and majority of the mechanical components made of a very pure 
Al-Si alloy. The cryostat has a J-type geometry to shield the crystal from 
radioactive impurities in the dewar. The passive shielding consisted of 4 cm 
of Roman-era lead and 3-10 cm of OFHC copper inside 15 cm of ordinary lead. To 
remove $^{222}$Rn gas, one of the main sources of the background, a special 
effort was made to minimize the free space near the detector. In addition, 
the passive shielding was enclosed in an aluminum box flushed with radon free 
air ($< 10$ mBq/m$^3$) delivered by a radon free factory installed in the Modane 
Underground Laboratory.

The electronics consisted of currently available spectrometric amplifiers and
an 8192 channel ADC. The energy calibration was adjusted to cover the energy 
range from 50 keV to 3.5 MeV, and the energy resolution was
2.0 keV for the 1332-keV line of $^{60}$Co. The electronics were stable during
the experiment due to the constant conditions in the laboratory (temperature 
of $\approx 23$ $^\circ$C, hygrometric degree of $\approx 50$\%).  A daily check
 on the apparatus assured that the counting rate was statistically constant.

\begin{figure}[h]
\begin{center}
\includegraphics[width=10.5cm]{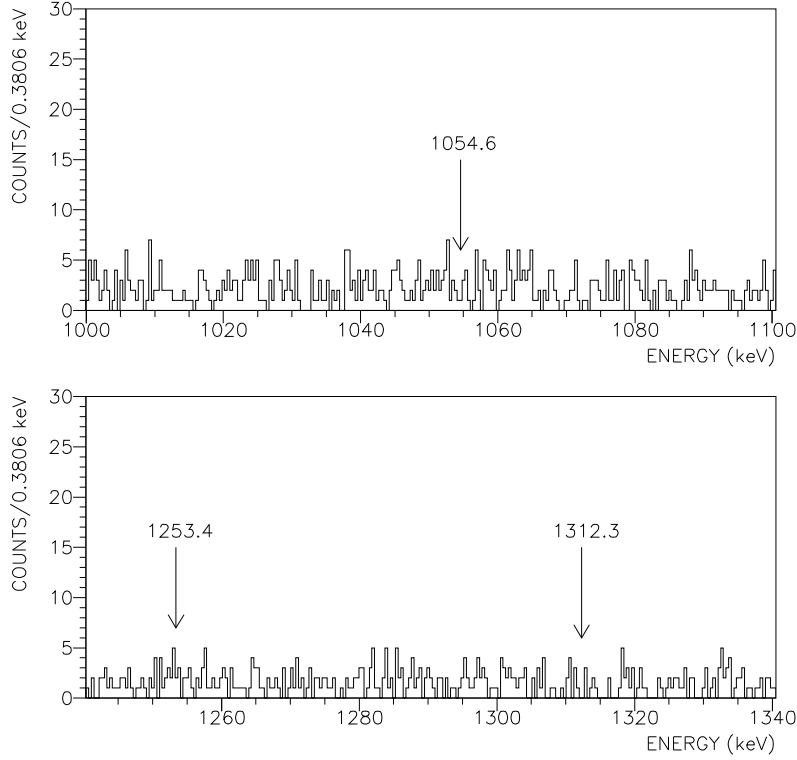}
\caption{Energy spectrum with 3975.3 g of natural Sn for 2385.4 h of measurement 
in the ranges investigated  ([1000-1100] and [1240-1340] keV).}  
\label{fig_2}
\end{center}
\end{figure}

\begin{figure}[h]
\begin{center}
\includegraphics[width=10.5cm]{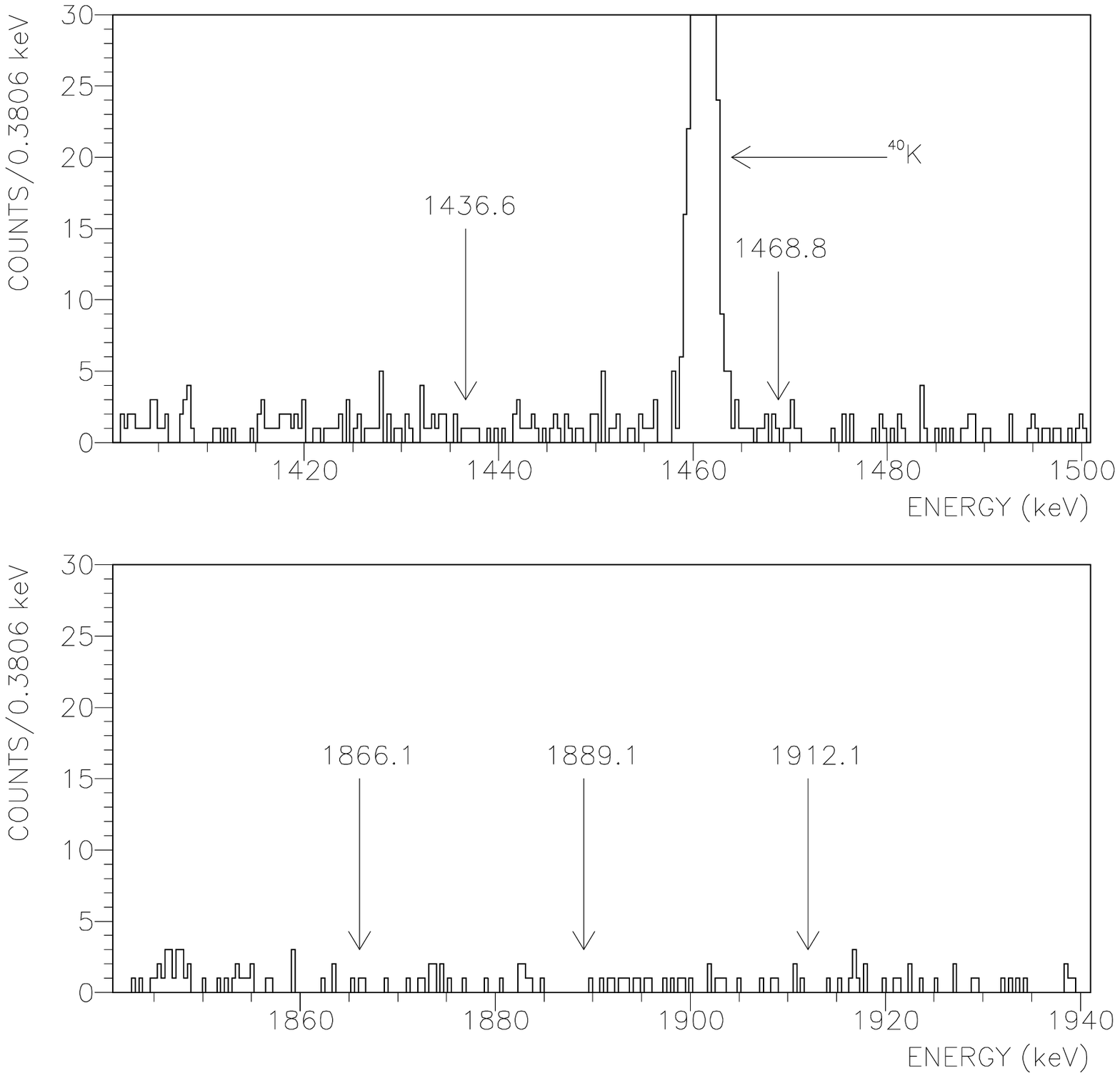}
\caption{Energy spectrum with 3975.3 g of natural Sn for 2385.4 h of measurement 
in the ranges investigated  ([1400-1500]  and [1850-1940] keV).}  
\label{fig_3}
\end{center}
\end{figure}

The first portion of natural tin (2543.2 g) was placed in a derlin Marinelli 
box surrounding the HPGe detector. The second portion (1432.1 g) was  placed 
 on the endcap 
of the HPGe detector. The total mass of tin was 3975.3 g, 240.24 g 
was $^{124}$Sn (natural abundance is 5.79\%) and 36.35 g was 
$^{112}$Sn (natural abundance is 0.97\%). The duration of the measurement 
was 2385.43 hours.

The natural tin sample was found to have a small quantity of $^{40}$K 
($(3.8\pm 0.6)$ mBq/kg) and also cosmogenic and natural tin radioactivities, 
i.e. the 158-keV decreasing peak of $^{117\rm m}$Sn (13.76 d), 
the 391-keV decreasing peak of $^{113}$Sn (115.09 d),
$(46\pm 9)\ \mu$Bq/kg of  
$^{126}$Sn ($10^5$ y), 
and $(159\pm 48)\ \mu$Bq/kg of $^{125}$Sb (2.75856 y). The natural 
radioactivities had only limits which were 
$<94\ \mu$Bq/kg of $^{228}$Ac, $<95\ \mu$Bq/kg of $^{226}$Ra,
$<37\ \mu$Bq/kg of $^{137}$Cs, and $<20\ \mu$Bq/kg of $^{60}$Co.

The search for different $\beta^+$EC and ECEC processes in $^{112}$Sn and 
$\beta^-\beta^-$ decay of $^{124}$Sn to the excited states of $^{124}$Te 
were carried out using the germanium detector to look for $\gamma$-ray lines
 corresponding to these processes. Gamma-ray spectra of selected energy ranges
 are shown in Fig. 1-3. These spectra correspond to regions-of-interest for 
the different decay modes of $^{112}$Sn and $^{124}$Sn.

\section{Search for $\beta^+$EC and ECEC processes in $^{112}$Sn}

The decay scheme for the triplet $^{112}$Cd-$^{112}$In-$^{112}$Sn is 
shown in Fig. 4 \cite{SIN96,TAB98}. 
\begin{figure}
\begin{center}
\includegraphics[width=10.5cm]{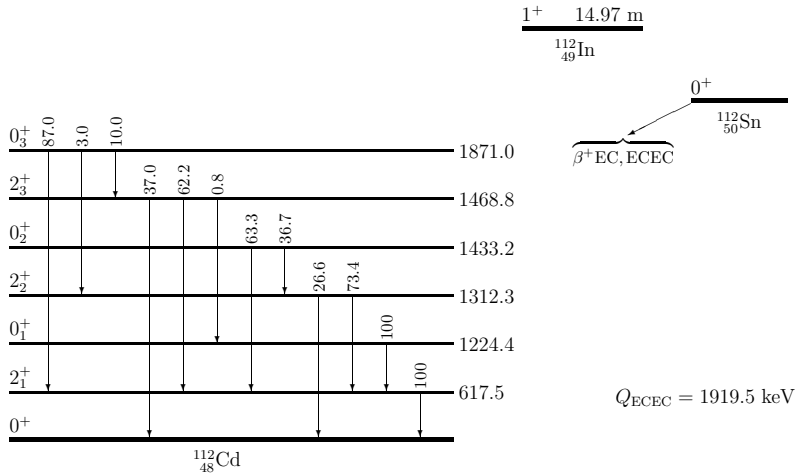}
\caption{Decay scheme of $^{112}$Sn. Only the investigated levels associated with 
$\gamma$-rays are shown. Transition probabilities are given in percents.}  
\label{fig_4}
\end{center}
\end{figure}
The $\Delta {\rm M}$ (difference of parent and daughter atomic masses) 
value of the transition is $1919.5. \pm 4.8$ keV \cite{AUD03} and the natural 
abundance of $^{112}$Sn is 0.97\%. 
The following decay processes are possible:

\begin{equation}
e^-_b + (A,Z) \rightarrow (A,Z-2) + e^{+} + X   \hspace{2cm}  (\beta^+EC; 0\nu)   
\end{equation}

\begin{equation}
e^-_b + (A,Z) \rightarrow (A,Z-2) + e^{+} + 2\nu + X   \hspace{1cm}  (\beta^+EC; 2\nu)   
\end{equation}

\begin{equation}
2e^-_b + (A,Z) \rightarrow (A,Z-2) + 2X   \hspace{2.5cm}  (ECEC; 0\nu)   
\end{equation}

\begin{equation}
2e^-_b + (A,Z) \rightarrow (A,Z-2) + 2\nu + 2X   \hspace{1.5cm}  (ECEC; 2\nu)   
\end{equation}
where e$_b$ is an atomic electron and X represents X-rays or Auger electrons.
Introduced here is the notation 
$Q'$ which is the effective $Q$-value defined as 
$Q'=\Delta {\rm M} - \epsilon_1-\epsilon_2$ 
for the ECEC transition and $Q'=\Delta {\rm M} - \epsilon_1- 2m_ec^2$ for the 
$\beta^+$EC process; $\epsilon_i$ is the electron binding energy of a daughter 
nuclide. For $^{112}$Cd, $\epsilon$ is equal to 26.7 keV for the K shell and 
4.01 keV, 3.72 keV and 3.54 keV for the L shell (2s, 2p$_{1/2}$ and 
2p$_{3/2}$ levels). In the case  of the L shell the resolution of the HPGe 
detector prohibits separation of the lines so we center the study on the 3.72 
keV line.

Investigations were made of the $\beta^+$EC transitions to the ground and the 2$^+_1$ 
excited states.
Additionally, the ECEC transitions to the ground state and six excited states 
(2$^+_1$, 0$^+_1$, 2$^+_2$, 0$^+_2$, 2$^+_3$ and 0$^+_3$) were investigated.

\subsection{ECEC transitions}  

The ECEC$(0\nu + 2\nu)$ transition to the excited states of $^{112}$Cd is 
accompanied with $\gamma$-quanta with different energies (see decay scheme in 
Fig. 4). These $\gamma$-quanta were used in the search. 
The approach is not sensitive to ECEC$(2\nu)$ to the ground state
 because X-rays are absorbed in the sample and cannot reach the sensitive 
volume of the HPGe detector.

The ECEC$(0\nu)$ transition to the ground state of the daughter nuclei was 
considered for three different electron capture cases:

1) Two electrons are captured from the L shell. In this case, $Q'$ is 
equal to $1912.1 \pm 4.8$ keV and the transition is accompanied by a 
bremsstrahlung $\gamma$-quantum with an energy $\sim$ 1912.1 keV.

2) One electron is captured from the K shell, another from the L shell. 
In this case, $Q'$ is equal to $1889.1 \pm 4.8$ keV and the transition is 
accompanied by a bremsstrahlung $\gamma$-quantum with an energy $\sim$ 
1889.1 keV.  

3) Two electrons are captured from the K shell. In this case, $Q'$ is equal
 to $1866.1 \pm 4.8$ keV and the transition is accompanied by 
$\gamma$-quantum with an energy $\sim$ 1866.1 keV. In fact this transition is 
strongly suppressed (forbidden) because of momentum conservation. So in this 
case the more probable outcome is the emission of $e^+e^-$ pair \cite{DOI93}
 which gives two annihilation $\gamma$-quanta with an energy of 511 keV.

The Bayesian approach \cite{PDG04} was used to estimate limits on transitions 
of $^{112}$Sn to the ground and excited states of $^{112}$Cd. To construct the 
likelihood function, every bin of the spectrum is assumed to have a Poisson 
distribution with its mean $\mu_i$ and the number of events equal to the 
content of the $i$th bin. The mean can be written in the general form,

\begin{equation}
\mu_i = N\sum_{m} {\varepsilon_m a_{mi}} + \sum_{k}
{P_k a_{ki}} + b_i .
\end{equation}

The first term in (5) describes the contribution of the investigated process 
that may have a few $\gamma$-lines contributing appreciably to the $i$th bin. 
The parameter $N$ is the number of decays, $\varepsilon_m$ is the detection 
efficiency of the $m$th $\gamma$-line and $a_{mi}$ is the 
contribution of the $m$th line to the $i$th bin. For low-background measurements a 
$\gamma$-line may be taken to have a gaussian shape. The second term gives 
contributions of background $\gamma$-lines. Here $P_k$ is the area of the 
$k$th $\gamma$-line and $a_{ki}$ is its contribution to the $i$th bin. 
The third term represents the so-called ``continuous background'' 
($b_i$), which has been selected as a straight-line fit after rejecting all peaks 
in the region-of-interest. We have selected this region as the peak to be 
investigated $\pm$ 30 standard deviations ($\approx$ 20 keV). The likelihood 
function is the product of probabilities for selected bins.  
Normalizing over the parameter $N$ gives the probability density 
function for $N$, which is used to calculate limits for $N$.  To take into 
account errors in the $\gamma$-line shape parameters, peak areas, and other 
factors, one should multiply the likelihood function by the error probability 
distributions for these values and integrate, to provide the average 
probability density function for $N$.

In the case of the ECEC$(0\nu)$ transition to the ground state of $^{112}$Cd 
there is a large uncertainty in the energy of the bremsstrahlung 
$\gamma$-quantum because of a poor accuracy in $\Delta M$ ($\pm$ 4.8 keV). 
Thus the position of the peak was varied in the region of the 
uncertainty and the most 
conservative value of the limit for the half-life was selected.       

The photon detection efficiency for each investigated process has been 
computed with the CERN 
Monte Carlo code GEANT 3.21. % \cite{GEANT21}.
Special calibration measurements with radioactive sources and powders 
containing well-known 
$^{226}$Ra activities confirmed that the accuracy of these efficiencies is 
about 10\%.

The final results are presented in Table 1. In the 4-th column are the best 
previous experimental results 
from Ref. \cite{KIM07,DAW07} for comparison. In the last column, 
the theoretical estimations for 
ECEC(2$\nu$) transitions obtained under the assumption of single intermediate 
nuclear state dominance 
are also presented \cite{DOM05}.

Concerning the ECEC$(0\nu)$ processes, the plan is to observe a resonant 
transition to the 1871.0 keV excited state of $^{112}$Cd. In this case we look 
for two peaks, at 617.5 keV and 1253.4 keV. In fact, 
the experimental spectrum has some excess of events in the range of 1253.4 keV, 
$\sim 2.3 \sigma$ above the continuous background. At the same time there 
are no extra events in the energy range of 617.5 keV. 
The total 
effect (taking into account both peaks) is only $1.5 \sigma$. If all the extra 
events are connected with the ECEC$(0\nu)$ 
transition of $^{112}$Sn to the 1871.0 keV excited state of $^{112}$Cd then the 
corresponding half-life value for this process is at the level of
$\sim 10^{20}$ y. The conservative approach gives the limit 
$T_{1/2} > 0.92\times 10^{20}$ y at the 90\% C.L. 

\subsection{$\beta^+$EC transitions}

The $\beta^+$EC$(0\nu + 2\nu)$ transition to the ground state is accompanied 
by two annihilation $\gamma$-quanta with an energy of 511 keV. These 
$\gamma$-quantum were used to search for this transition. In the case of the 
$\beta^+$EC$(0\nu + 2\nu)$ transition to the 2$^+_1$ excited state the 
617.4 keV $\gamma$-quanta was also detected. To obtain limits on these 
transitions the analysis described in Section 3.1 was used. Again the 
photon detection efficiencies for each investigated process was computed 
with the CERN Monte Carlo code GEANT 3.21. % \cite{GEANT21} 
and are presented in Table 1. 
In the last two columns are the best previous results and theoretical predictions
for comparison.
 
\begin{table}
\caption{\label{arttype1}The experimental limits and theoretical predictions 
for the $\beta^+$EC and ECEC processes in 
$^{112}$Sn. $^{*)}$ For transition with irradiation 
of $e^+e^-$ pair - see text.}
\begin{tabular*}{\textwidth}{lllll}
\hline

%\br
Transition & Energy of $\gamma$-rays, & \multicolumn{2}{c}{$T_{1/2}^{exp}$, 
$10^{20}$ y (C.L. 90\%)} & $T_{1/2}^{th}(2\nu)$, y \\
\cline{3-4}
 & keV (Efficiency) & Present & Previous & \cite{DOM05} \\
 &  & work & works \\
\hline
%\mr
$\beta^+$EC$(0\nu + 2\nu)$; g.s. & 511.0 (5.05 \%) & 0.12 & 0.0091  \cite{KIM07} & $3.8\times 10^{24}$ \\
 &  &  & 0.041  \cite{DAW07}\\
$\beta^+$EC$(0\nu + 2\nu)$; 2$^+_1$ & 617.5 (1.85 \%) & 0.94 & 0.014  \cite{DAW07} & $2.3\times 10^{32}$\\
\\
ECEC$(0\nu)$ ${\rm L}^1{\rm L}^2$; g.s. & 1912.1 (1.44 \%) & 1.3 & - \\
ECEC$(0\nu)$ ${\rm K}^1{\rm L}^2$; g.s. & 1889.1 (1.45 \%) & 1.8 & 0.0099  \cite{DAW07}\\
ECEC$(0\nu)$ ${\rm K}^1{\rm K}^2$; g.s. & 1866.1 (1.47 \%) & 1.3 & - \\
 & 511.0 (5.05 \%) & 0.12$^{*)}$ & -\\
\\
ECEC$(0\nu)$; 2$^+_1$ & 617.5 (2.11 \%) & 1.1 & 0.014  \cite{DAW07} &  \\
ECEC$(0\nu)$; 0$^+_1$ & 606.9 (1.90 \%) & 1.2 & 0.014  \cite{DAW07} &  \\
                      & 617.5 (1.87 \%) \\ 
ECEC$(0\nu)$; 2$^+_2$ & 617.5 (1.37 \%)  & 0.89 & 0.014  \cite{DAW07} &  \\
                      & 1312.3 (0.46 \%) \\          
ECEC$(0\nu)$; 0$^+_2$ & 617.5 (1.73 \%)  & 1.6 & 0.014  \cite{DAW07} &  \\
                      & 815.8 (1.08 \%) \\
ECEC$(0\nu)$; 2$^+_3$ & 617.5 (1.20 \%)  & 0.93 & 0.014  \cite{DAW07} &  \\
                      & 851.1 (1.06 \%) \\
                      & 1468.8 (0.58 \%) \\  
ECEC$(0\nu)$; 0$^+_3$ & 617.5 (2.00 \%)  & 0.92 & 0.016  \cite{DAW07} &  \\
                      & 1253.4 (1.39 \%) \\
\\
ECEC$(2\nu)$; 2$^+_1$ & 617.5 (2.41 \%) & 1.2 & 0.014  \cite{DAW07} & $4.9\times 10^{28}$\\
ECEC$(2\nu)$; 0$^+_1$ & 606.9 (2.11 \%) & 1.4 & 0.014  \cite{DAW07} & $7.4\times 10^{24}$\\
                      & 617.5 (2.09 \%) \\  
ECEC$(2\nu)$; 2$^+_2$ & 617.5 (1.55 \%)  & 1.0 & 0.014  \cite{DAW07} & $1.9\times 10^{32}$\\
                      & 1312.3 (0.51 \%) \\

\hline
\end{tabular*}
\end{table}

%Table 1 continued
\addtocounter{table}{-1}
\begin{table}
\caption{Continued.}
\bigskip

\begin{tabular*}{\textwidth}{lllll}

\hline
Transition & Energy of $\gamma$-rays, & \multicolumn{2}{c}{$T_{1/2}^{exp}$, 
$10^{20}$ y (C.L. 90\%)} & $T_{1/2}^{th}(2\nu)$, y \cite{DOM05}\\
\cline{3-4}
 & keV (Efficiency) & Present & Previous & \\
 &  & work & works \\
\hline

ECEC$(2\nu)$; 0$^+_2$ & 617.5 (1.88 \%)  & 1.8 & 0.014  \cite{DAW07} & - \\
                      & 815.8 (1.21 \%) \\     
ECEC$(2\nu)$; 2$^+_3$ & 617.5 (1.32 \%)  & 1.0 & 0.014  \cite{DAW07} & $6.2\times 10^{31}$\\
                      & 851.1 (1.19 \%) \\
                      & 1468.8 (0.64 \%) \\  
ECEC$(2\nu)$; 0$^+_3$ & 617.5 (2.00 \%)  & 0.92 & 0.016  \cite{DAW07} & $5.4\times 10^{34}$\\
                      & 1253.4 (1.39 \%) \\

\hline
\end{tabular*}
\end{table}

\section{Search for $\beta\beta$ decay of $^{124}$Sn to the excited states 
of $^{124}$Te}

The decay scheme for the triplet $^{124}$Sn-$^{124}$Sb-$^{124}$Te is shown 
in Fig. 5 \cite{TAB98,IIM97,LED78}. The $\Delta {\rm M}$ in this case is 2287.7 $\pm$
 2.1 keV \cite{AUD03} . The following processes are possible:

\begin{figure}
\begin{center}
\includegraphics[width=10.5cm]{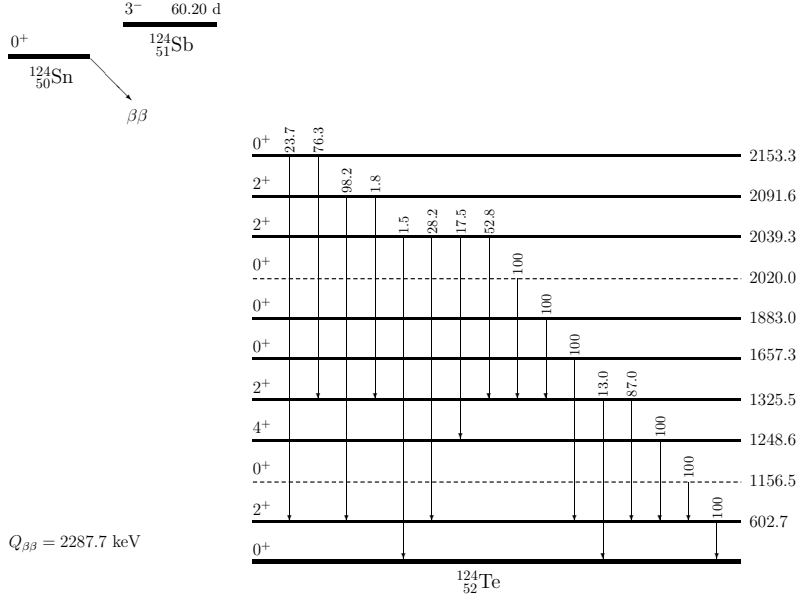}
\caption{Decay scheme of $^{124}$Sn are taken from \cite{IIM97} and two levels 
(1156.5 keV and 2020.0 keV) are added from \cite{LED78}. The investigated levels are
associated with the $\gamma$-rays shown. Transition probabilities are given in percents.}  
\label{fig_5}
\end{center}
\end{figure}

\begin{equation}
(A,Z) \rightarrow (A,Z+2) + 2e^{-} + 2\tilde \nu  \hspace{2cm}  (2\nu\beta\beta)   
\end{equation}

\begin{equation}
(A,Z) \rightarrow (A,Z+2) + 2e^{-}  \hspace{2cm}  (0\nu\beta\beta)   
\end{equation}

\begin{equation}
(A,Z) \rightarrow (A,Z+2) + 2e^{-} + \chi^0   \hspace{2cm}  (0\nu\chi^0\beta\beta)   
\end{equation}

We will consider only transitions to the excited states of $^{124}$Te. 
The most intensive $\gamma$-rays from the decay scheme
(Fig. 5) were used for analysis. The detection efficiencies for photopeaks 
corresponding to specified $\gamma$-quanta are given in Table~2.  
The efficiencies were again computed with
the CERN Monte Carlo code GEANT 3.21. % \cite{GEANT21}. 

The spectra for specified
energy ranges are shown in Fig. 1-3.
There are no statistically significant peaks at the 
indicated energies. 
The lower half-life limits reported in Table 2 have been calculated using
the likelihood function described in Section 3.1.  
Available data on $\beta\beta$ decay of $^{124}$Sn from other 
experimental works and theoretical estimates 
are presented in Table 2.

Our limits on $\beta\beta$ transitions of $^{124}$Sn to the excited states
of daughter nucleus are valid for the $0\nu$, $2\nu$ modes and all types 
of decay with Majoron emission ($0\nu\chi^0$).

\begin{table}
\caption{Theoretical and experimental results for $(0\nu+2\nu+0\nu\chi^0)\beta\beta$ decay of $^{124}$Sn 
to the excited states of $^{124}$Te. Limits are given at the 90\% C.L.}
%\footnotesize\rm
%\begin{tabular*}{\textwidth}{lllll}
\begin{tabular}{ccccc}
\hline
Excited state, & Energy of $\gamma$-rays, & $(T^{2\nu}_{1/2})^{th}$, y \cite{AUN96} &
\multicolumn{2}{c}{$(T^{0\nu+2\nu+0\nu\chi^0}_{1/2})_{exp}$, $10^{21}$ y} \\ 
 \cline{4-5}
 (keV) & keV (Efficiency) & & Present work & Previous works \\ \hline
\hline
$2^+_1 (602.7) $  & 602.7 (2.40\%) & $6.5\cdot 10^{26}$ & $ > 0.91 $ & $ > 0.0031$ \cite{DAW07}\\
                  &                &                    &           & $ > 0.0023$ \cite{KIM07}\\
                  &                &                    &           & $ > 0.033$ \cite{SMO85}\\
$0^+_1 (1156.5)$  & 553.8 (2.20\%) & $2.7\cdot 10^{21}$ & $ > 1.1 $ & $ > 0.0077$ \cite{DAW07}\\
                  & 602.7 (2.15\%) &                    &           & $ > 0.0067$ \cite{KIM07}\\
$2^+_2 (1325.5)$  & 602.7 (1.86\%) & $1.7\cdot 10^{27}$ & $ > 0.94 $ & $ > 0.0044$ \cite{DAW07}\\
                  & 722.8 (1.76\%) &                    &           & $ > 0.0079$ \cite{KIM07}\\
$0^+_2 (1657.3)$  & 602.7 (2.14\%) & $ - $              & $ > 1.2 $ & $ > 0.0079$ \cite{DAW07}\\ 
                  & 1054.6 (1.72\%) \\  
$0^+_3 (1882.98)$ & 557.5 (1.96\%) & $ - $              & $ > 1.2 $ & $ - $ \\
                  & 602.7 (1.65\%) \\ 
                  & 722.8 (1.55\%) \\  
$0^+_4 (2020.0)$  & 602.7 (1.64\%) & $ - $              & $ > 0.82 $ & $ > 0.0044$ \cite{DAW07}\\ 
                  & 694.5 (1.81\%) \\ 
                  & 722.8 (1.53\%) \\  
$2^+_3 (2039.3)$  & 602.7 (1.82\%) & $ - $              & $ > 0.86 $ & $ > 0.0044$ \cite{DAW07}\\ 
                  & 722.8 (0.82\%) \\  
                  & 1436.6 (0.46\%) \\ 
$2^+_4 (2091.6)$  & 602.7 (2.13\%) & $ - $              & $ > 0.96 $ & $ > 0.0031$ \cite{DAW07}\\ 
                  & 1488.9 (1.43\%) \\ 
$0^+_5 (2153.3)$  & 602.7 (1.78\%) & $ - $              & $ > 0.95 $ & $ - $ \\ 
                  & 722.8 (1.18\%) \\  
                  & 1550.6 (0.37\%) \\ 
\hline
%\rule[-2mm]{0mm}{5mm}  
%\rule[-2mm]{0mm}{5mm}  
\end{tabular}
\end{table}

\section{Discussion} 

Limits obtained for the $\beta^+$EC and ECEC processes in $^{112}$Sn are on the level of 
$\sim (0.1-2)\times 10^{20}$ y or $\sim$ 10-200 times better 
than the best previous results \cite{KIM07,DAW07} (see Table 1). 
As one can see from Table 1 the theoretical predictions for $2\nu$ transitions are much 
higher than the measured limits. The sensitivity of such experiments can still be increased with the 
experimental possibilities being the following:  
 \\ 1) Given 4 kg of enriched $^{112}$Sn in the setup described in   
Section 2, the sensitivity after one year of measurement will be $\sim 3 \cdot 10^{22}$ y. \\ 
2) With
200 kg of enriched $^{112}$Sn using an installation such as GERDA \cite{ABT04} or MAJORANA 
\cite{MAJ03,AAL05} where 500-1000 kg of low-background HPGe detectors are planned.\footnote{Imaging 
around each of the $\sim$ 200 HPGe crystals can be $\sim$ 1 kg of very pure $^{112}$Sn. Both 
$^{76}$Ge and $^{112}$Sn will be investigated at the same time.} The 
sensitivity after 10 years of measurement may reach $\sim 10^{26}$ y. Thus there is 
a chance of detecting the $\beta^+$EC(2$\nu$) transition of $^{112}$Sn to the ground state 
and ECEC(2$\nu$) transition to the $0^+_1$ excited state (see theoretical predictions in Table 1). 

In the case of the ECEC(0$\nu$) transition to the $0^+_3$ (1871.0 keV) excited state of $^{112}$Cd a 
very small excess  of 
events ($\sim 1.5\sigma$) was detected. The effect is weak and only a limit can be 
extracted in this case. So the 
search for this process continues into the future. 
Note that the ECEC(2$\nu$) transition to the $0^+_3$ excited state is strongly suppressed because of 
the very small phase space volume. In contrast, the probability of the 0$\nu$ transition should be 
strongly enhanced if the resonance condition is realized. In Ref. \cite{BER83,SUJ04} the "increasing factor" 
was estimated as $\sim 10^6$ and can be even higher. Then if the "positive" effect is observed in  
future experiments it is the ECEC(0$\nu$)
process. This will mean that lepton number is violated and the neutrino is a Majorana particle. To extract the
$\left<m_\nu\right>$ value one has to know the nuclear matrix element for this transition and therefore the exact 
value of $\Delta {\rm M}$ (see \cite{BER83,SUJ04}). The necessary accuracy for $\Delta {\rm M}$ 
is better than 1 keV and this is  
a realistic task (in Ref. \cite{FRE05} the possibility of measurements with accuracy $\sim 200$ eV is discussed).

Two of different descriptions for the resonance were discussed in the past. In Ref. \cite{BER83} the 
resonance condition is realized when $Q'$ is close to zero. They treat the process as (1S,1S) double 
electron capture and $Q'$ is equal to $-4.9 \pm 4.8$ keV (1$\sigma$ error). 
Thus there is a probability that $Q'$ is less than 1 keV. 
In this case one has a few 
daughter-nucleus $\gamma$ rays (see scheme in fig.1) and two Cd K X-rays, one of which may have 
its energy shifted by the mismatch in energies between the parent atom and the almost degenerate virtual 
daughter state. In Ref. \cite{SUJ04,LUK06}
the decay is treated as (1S,2P) double electron capture with irradiation of an internal bremsstrahlung photon. 
The $Q'$ value (energy of the bremsstrahlung photon) is $18.1 \pm4.8$ keV. The resonance condition for the transition 
is realized when $E_{brems}=Q_{res}=\mid E(1S,Z-2)-E(2P,Z-2)\mid $, i.e. when the 
bremsstrahlung photon energy becomes comparable to the $2P-1S$ atomic level difference in the final 
atom (23 keV)\footnote{The same effect was theoretically predicted and then experimentally confirmed for 
single electron capture (see discussion in \cite{LUK06}) }.
Anticipated, taking into account uncertainties in the $Q'$ value, is that the real $Q'$ value is equal 
to 23 keV with an accuracy better then 1 keV and the resonance condition is realized. 
There are a few 
daughter-nucleus $\gamma$ rays (see scheme in fig.1), one Cd K X-ray and bremsstrahlung photon with energy 
$\sim K_{\alpha}$. 
The bremsstrahlung photon may have 
its energy shifted by the mismatch in energy between the parent atom and the almost degenerate virtual 
daughter state.

Finally, both approaches predict the same experimental signature for this transition and need to know with better accuracy 
the value of $\Delta {\rm M}$ to be sure that the resonance condition is really valid. New 
theoretical investigations of this transition are needed. 
   
Limits obtained for $\beta\beta$ decay of $^{124}$Sn to the excited state of $^{124}$Te are on the 
level $\sim (0.8-1.2)\times 10^{21}$ y 
or $\sim$ 30-300 times better 
than the best previous results \cite{KIM07,DAW07,SMO85} (see Table 2). 
One can see from Table 2 the theoretical prediction for $2\nu$ transitions to the $0^+_1$ excited state 
is $2.7\cdot 10^{21}$ y and only 
$\sim 2.5$ times higher than the obtained limit. Therefore there is a chance to detect this decay in the future 
with more
sensitive experiments. The experimental possibilities here are approximately the same as for $^{112}$Sn (see above):  
 \\ 1) With 4 kg of enriched $^{124}$Sn in the experiment described in   
Section 2, the sensitivity after one year of measurement will be $\sim 3 \cdot 10^{22}$ y. \\ 
2) With
200 kg of enriched $^{124}$Sn using an installation such as GERDA \cite{ABT04} or MAJORANA 
\cite{MAJ03,AAL05} where 500-1000 kg of low-background HPGe detectors are planned. The 
sensitivity after 10 years of measurement may reach $\sim 10^{26}$ y.

\section{Conclusion}

New limits on $\beta^+$EC and ECEC processes in $^{112}$Sn and on $\beta^-\beta^-$ decay 
of $^{124}$Sn to the excited states of $^{124}$Te have been obtained using 
a 380 cm$^3$ HPGe detector and an external source consisting of natural tin. 
In addition, it has been demonstrated that, in the future larger-scale experiments, the sensitivity to the ECEC($0\nu$) 
processes for $^{112}$Sn can reach the order of $10^{26}$ y. Under resonant 
conditions this decay will be competitive with $0\nu\beta\beta$ decay. The same sensitivity can be reached for 
double beta decay of $^{124}$Sn to the excited states of $^{124}$Te.

\section*{Acknowledgement}
The authors would like to thank the Modane Underground Laboratory staff for their technical 
assistance in running the experiment. We are very thankful to Prof. S. Sutton for his useful remarks. 
Portions of this work were supported by a grant from RFBR 
(no 06-02-72553) and State contract 02.516.11.6099. 
This work was supported by Russian Federal Agency for Atomic Energy.

 --------------------------------------------------------------

\end{document}